\documentclass{article} 

\usepackage{spconf}
\usepackage{graphicx}
\usepackage{subcaption}
\usepackage{mwe}
\usepackage{adjustbox}
\usepackage{authblk}

\title{Ink flow patterns in multi color inkjet images and their impact on graininess noise}
\name{Qiulin Chen$^{\star}$ \qquad Palghat Ramesh$^{\dagger}$ \qquad Chu-heng Liu$^{\dagger}$ \qquad Jan P. Allebach$^{\star}$}
 \address{$^{\star}$ Purdue University, ECE Department, West Lafayette, IN 47907  \\ $^{\dagger}$ Advanced Technology Group, Xerox Labs, Webster, NY 14580}

\begin{document}
\maketitle
\begin{abstract}
Graininess noise is a common artifact in inkjet printing. While current inkjet printing technologies attempt to control graininess in single color images, the results are often less than optimal for multi-color images. This is due to fluidic interactions between inks of different colors. This paper will describe a color decomposition methodology that can be used to study ink flow patterns in multi-color inkjet printed images at a microscopic scale. This technique is used to decompose multi-color images into several independent color components. The ink patterns in these components is analyzed to relate them to visually perceptible graininess noise.
\end{abstract}

\section{Introduction}

\begin{figure}[h!]
    \centering
    \captionsetup{justification=centering}
        \begin{subfigure}[b]{0.234\textwidth}
            \centering
            \includegraphics[trim={0 35mm 0 0},clip,width=\textwidth]{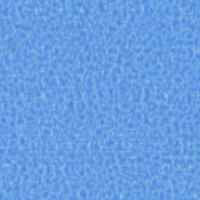}
            \caption[Network2]%
            {{\small Cyan only (60\%)}}    
            \label{fig:Epson Cyan only}
        \end{subfigure}
        \hfill
        \begin{subfigure}[b]{0.234\textwidth}
            \centering 
            \includegraphics[trim={0 35mm 0 0},clip,width=\textwidth]{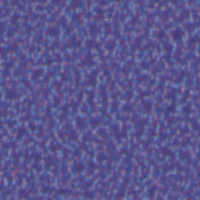}
            \caption[]%
            {{\small Cyan (60\%), Magenta (45\%)}}    
            \label{fig:Epson Cyan+Magenta}
        \end{subfigure}
        \vskip\baselineskip
        \begin{subfigure}[b]{0.234\textwidth}   
            \centering 
            \includegraphics[trim={0 35mm 0 0},clip,width=\textwidth]{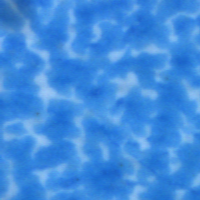}
            \caption[]%
            {{\small Cyan only (60\%)}}    
            \label{fig:PiasII Cyan only}
        \end{subfigure}
        \hfill
        \begin{subfigure}[b]{0.234\textwidth}   
            \centering 
            \includegraphics[trim={0 35mm 0 0},clip,width=\textwidth]{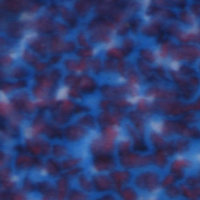}
            \caption[]%
            {{\small Cyan (60\%), Magenta (45\%)}}    
            \label{fig:PiasII Cyan+Magenta}
        \end{subfigure}
        \captionsetup{justification=raggedright}
        \caption{Example of graininess noise in inkjet printed pages. Images in first row are scanned with an EPSON A3 scanner at 1200 dpi. Images in second row are scanned with a $PIAS\, II$ scanner at 8000 $dpi$.}
        \label{fig:dataset zoom in}
\end{figure}

Human vision perceptible image noise is of great importance to the printing industry. Mottle (low frequency noise) and graininess (high frequency noise) are two kinds of image noise that is particularly relevant. In the ink-jet printing industry, much of the focus has been on controlling image noise (graininess and mottle) in single colors. But for multi-color images, image noise can be significantly worse. One potential reason is that, for single colors patches, the image noise is primarily dictated by ink-paper interactions . But in multi-color printing, the ink drops of a different color can overlay on the previous color ink layer. The resulting ink-ink interaction can create complex high frequency ink flow patterns that can manifest as visually perceptible image noise.  An example of human perceptible visual noise in single color and multi-color images is shown in Figures \ref{fig:Epson Cyan only} and \ref{fig:Epson Cyan+Magenta}, respectively. The corresponding microscopic ink flow patterns
in these images are shown in \ref{fig:PiasII Cyan only} and \ref{fig:PiasII Cyan+Magenta}, respectively.

Ink-jet printing is a complex physical process. There are
many sources of noise within a printer, such as variations in
drop size, drop velocity, drop direction, in addition to noises
associated with the paper. For instance, due to directionality
errors, ink drops jetted by the printhead may not land at the location
expected in the halftoned image. Also, the ink spreads
on the paper surface before stabilizing, a process that depends
on the drop size, drop velocity and the relative surface energies
of the ink and the paper surface it contacts. These introduce
further uncertainties in predicting the ink pattern distributions.
Finally, when printing with multiple colors, interaction
between different color ink drops, i.e. coalescence and
color mixing, will also introduce noise to the output. Consequently
it is almost impossible to accurately predict the output
of printed pages at a microscopic level.

Image noise in printing is typically viewed at two scales:
macroscopic and microscopic. The macroscopic scale noise,
i.e. macro-uniformity, includes image defects such as streaks,
bands, spots, mottle and chromatic variations.  Rasmussen and Dalal \cite{rasmussen2001measurement,dalal1998evaluating} discussed the effect of macro-uniformity on the overall image quality of hardcopy output from printer systems and introduced a high-level set of image quality attributes. Rasmussen also created simulated macro-uniformity defects and built subjective evaluations using quality ruler \cite{rasmussen2006iso}. Wang \textit{et al.} \cite{wang2013figure} built a learning based model to predict image macro-uniformity. More recently, researchers \cite{chen2019segmentation, huang2019cost, jing2013general, xiang2019blockwise, yan2015autonomous,zhang2019block} have tried to automatically detect specific defects instead of investigating the uniformity of whole printed pages. \cite{nguyen2014perceptual} discusses how to automatically rank printed pages \cite{nguyen2014perceptual}. In the microscopic scale, image graininess is measured as high frequency noise in the $1 \sim10 \; cycles/mm$ range. Dirk \textit{et al.} \cite{hertel2003one} designed a scanner-based method to measure
graininess versus optical density for digital photo print systems. Tse and Forrest \cite{tse2009towards} investigated the relation between calculated variance and human perceptible mottle and graininess. They  created a composite noise index (CNI) metric that correlated well with subjective assessment of image uniformity. While previous work on graininess has focused on measuring the noise level, relatively little is known about root causes of graininess, particularly in inkjet printing.


In this paper, multi-color images printed with cyan and magenta inks in an inkjet printer are analyzed to understand the role of ink-ink interactions on ink flow patterns at the microscopic level and relate them to visually perceptible graininess noise. The methodology, described in Section 2, is divided into two parts: (1) a model-based color segmentation framework to segment different color components from scanned multi-color printed pages, (2) a color reflectance model to estimate the contribution of each color component to the total reflectance of a multi-color patch. In section 3, the proposed approach is tested on a dataset comprising of patches with varying amounts of cyan and magenta inks, and experimental results to support our conclusion are discussed.


\section{Methodology}
\subsection{Color segmentation pipeline}
\begin{figure}[h!]
    \centering
    \includegraphics[width=0.48\textwidth]{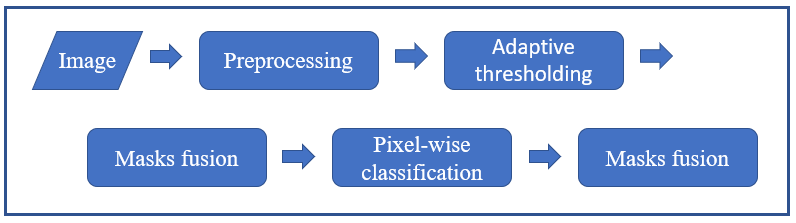}
    \caption{Color segmentation pipeline.}
    \label{fig:pipeline}
\end{figure}

Figure \ref{fig:pipeline} shows the pipeline for multi-color segmentation. First the scanned printed pages are preprocessed to increase contrast for different color channels (RGB). Then adaptive thresholding is applied in each color channels to achieve raw segmentation masks for each intermediate color component (e.g. cyan+white, magenta+white, cyan+black). Here white refers to regions with no ink, while black refers to regions of overlapping (coalesced) ink layers (e.g. cyan+magenta). Next, the results from different color channels are fused and fed into a pixel-wise classifier to further refine the segmentation results. The pixel-wise classifier includes Otsu's method \cite{otsu1979threshold} and K-nearest neighbour (KNN) clustering algorithm \cite{cover1967nearest}. After classification, the refined masks from the different color channels are fused to obtain the segmentation masks for the final color components (e.g. pure cyan, pure magenta, cyan+magenta overlap and white) .

To evaluate the performance of the color segmentation pipeline, the mean intersection over union (mIoU) and pixel wise accuracy are used as metrics. The mIoU calculates the mean value of the ratio between area of intersection and area of union.The mIoU between the segmentation masks in set $A$ and their corresponding ground truth masks in set $B$ is calculated as:
\begin{equation}
        mIoU = \frac{1}{|A|} \sum_{i=1}^{N}{\frac{A_i\cup B_i}{A_i\cap B_i}} \ , A_i \in A \ \textrm{for} \ \forall i \in [1,|A|]
\end{equation}

\subsection{Reflectance Model}

Human perceptible graininess noise is related to modulations in luminescence (L*) of an image in the 1-10 cycles/mm range. A reflectance model is used to relate color components from segmentation to the L* of the patch, thereby allowing us to investigate how different color components contribute to image graininess.  The $ Murray$-$Davis \; Model$ and $ Neugebauer \; Theory $ are used to develop a relation between reflectance of a color patch and individual reflectance from the different color components in it.  

$Murray$-$Davis \; Model$ assumes that the total reflectance of a paper surface covered with ink layers is equal to the weighted summation of reflectance of different areas. This model focuses on halftoned monochromatic prints, where there are only two types of regions, black regions with ink and white regions without ink.  $Neugebauer \; Theory$ extends the $Murray$-$Davis \; Model$ to predict reflectance of multiple color halftones.  Neugebauer represents halftoned color prints as a mosaic of eight colors (white, cyan, magenta, yellow, red, green, blue, and black), called the $ Neugebauer \ primaries $. 
\begin{equation} \label{eq:Neugebauer}
    R(\lambda)=\sum_{j=1}^{8}{a_jR_j(\lambda)}    \quad with \quad \sum_{i=1}^8{a_i=1}
\end{equation}
$R_j$ is the reflectance and $a_j$ the area coverage ratios of the $j^{th}$ color primary.

In the present work, Equation \ref{eq:Neugebauer} is modified for use on high resolution scans of printed pages instead of halftoned images, with color components from the segmentation pipeline as the primaries. Let $S_{ink}=\{pc,pm,w,o\}$ represents a label set for the four ink components, pure cyan, pure magenta, white and overlap, respectively.  Then,
\begin{equation} \label{eq:our model}
    R_{total} = \sum_{k\in S_{ink}}{a_kR_k} \quad with \; \sum_{k\in S_{ink}}{a_k = 1}
\end{equation}
In the equation above, $a_{pc}$, $a_{pm}$, $a_w$ and $a_o$ are area coverage ratios for different color components calculated from segmentation masks.

\begin{figure}[h]
    \centering
    \captionsetup{justification=centering}
     \includegraphics[width=0.4\textwidth]{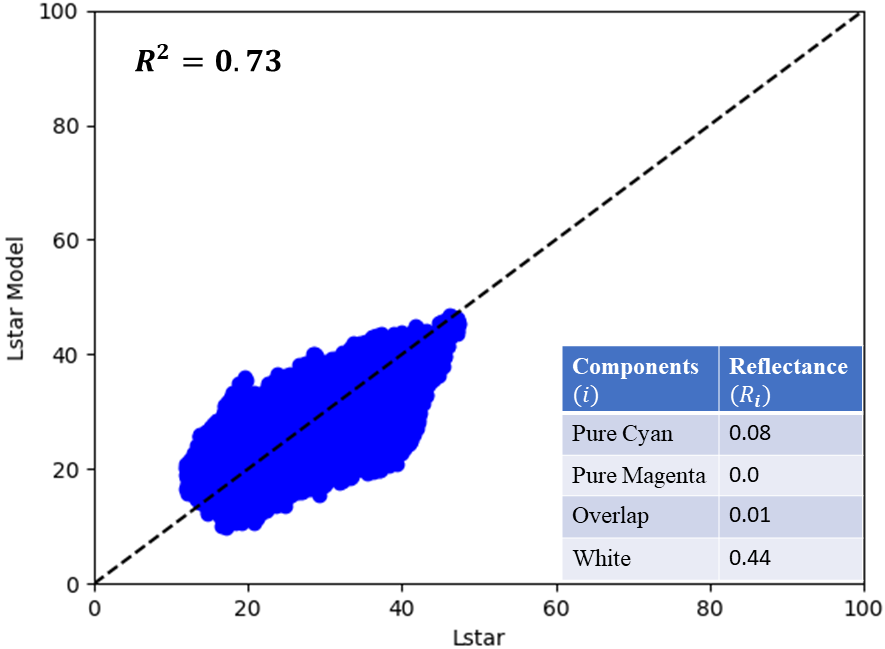}
     \caption{Linear regression fit for reflectance model.}
     \label{fig:L_star model}
\end{figure}

To estimate reflectance of different color components in Equation \ref{eq:our model}, the total reflectance of color patches, obtained using CIELAB color conversion formulae applied to measured L* of the patch \cite{berns2000principles}, is fitted to the area coverage ratios of the color components in $S_{ink}$ in color segmentation masks.  Figure \ref{fig:L_star model} shows the linear regression fit for gloss-coated paper. The fitted coefficients are the reflectances for the different color components.
The fitted reflectances of the color components  (shown in Figure \ref{fig:L_star model} inset) suggests that white (no ink) regions contribute the most to the overall reflectance of the patch (and thus its luminescence), while other components contribute far less.  This has implications to the root cause graininess, as discussed next. 

\section{Results and Discussion}
\subsection{Dataset}
\begin{figure}[h!]
    \centering
    \includegraphics[width=0.45\textwidth]{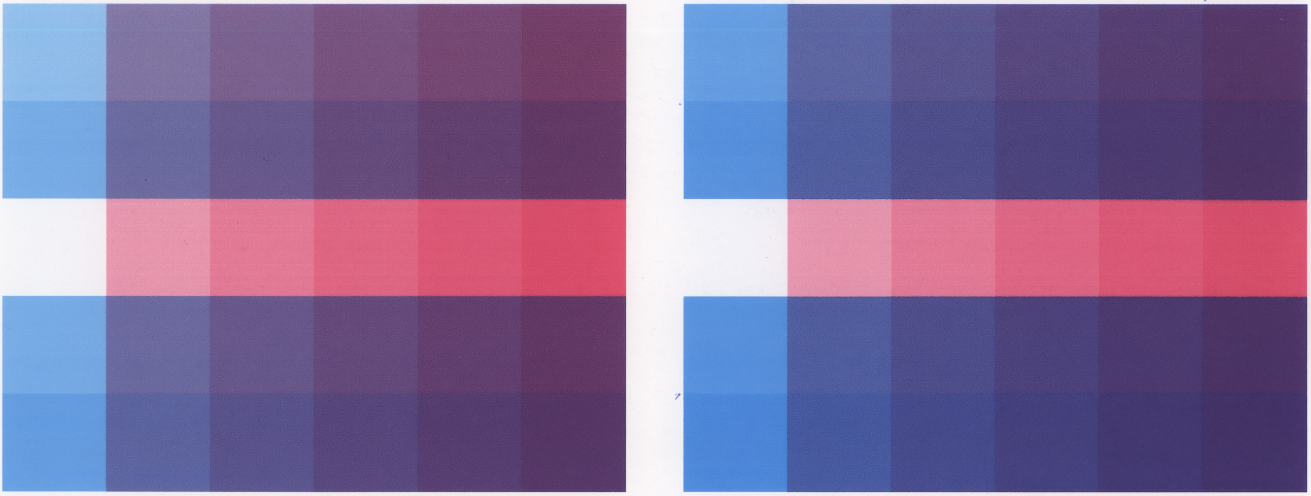}
    \caption{Test image used generate the dataset.}
    \label{fig:dataset}
\end{figure}

Color patches with varying amounts of cyan and magenta ink amounts are printed using an inkjet printer. Figure \ref{fig:dataset} shows the image used, where cyan and magenta ink levels are varied between 0\%, 30\%, 45\% 60\% 75\% and 90\%, allowing for both single color (pure cyan or pure magenta) and multi-color patches to be printed at different ink level combinations.
The prints are scanned using a $PIAS \; II$ scanner \cite{tse2007pias} in high resolution mode (8000 \textit{dpi}), with a field of view of 3.2 mm by 2.4 mm. Since the size of an ink drop on paper is about 30 microns, the scanned patch is sufficient for a representative sampling of the stochastic behavior of ink spreading and coalescence.  An example is shown in Figure \ref{fig:decomposition example a}.

\subsection{Color segmentation of multi-color images}
\begin{figure}[h]
    \centering
    \captionsetup{justification=centering}
        \begin{subfigure}[b]{0.234\textwidth}
            \centering
            \includegraphics[trim={0 30mm 0 0},clip, width=\textwidth]{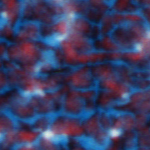}
            \caption[Network2]%
            {{\small Multi-color image}}
            \label{fig:decomposition example a}
        \end{subfigure}
        \hfill
        \begin{subfigure}[b]{0.234\textwidth}
            \centering 
            \includegraphics[trim={0 30mm 0 0},clip,width=\textwidth]{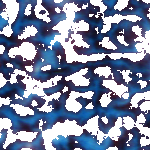}
            \caption[]%
            {{\small Cyan mask}}    
            \label{fig:decomposition example b}
        \end{subfigure}
        \begin{subfigure}[b]{0.234\textwidth}   
            \centering 
            \includegraphics[trim={0 30mm 0 0},clip,width=\textwidth]{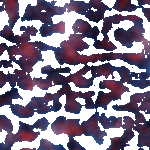}
            \caption[]%
            {{\small Magenta mask}} 
            \label{fig:decomposition example c}
        \end{subfigure}
        \hfill
        \begin{subfigure}[b]{0.234\textwidth}   
            \centering 
            \includegraphics[trim={0 30mm 0 0},clip,width=\textwidth]{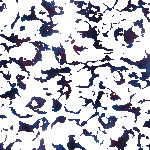}
            \caption[]%
            {{\small Overlap mask}}
            \label{fig:decomposition example d}
        \end{subfigure}
        \caption{An example of color segmentation results.}
        \label{fig:decomposition example}
\end{figure}

Figure \ref{fig:decomposition example} shows an example by applying the color segmentation method on a scanned image. It shows a scanned image printed with 60\% cyan ink amount and 60\% magenta ink and segmentation masks for cyan, magenta and overlap regions, respectively.

Since it is hard to do human annotate on the original dataset, the segmentation pipeline is tested on synthetic ground truth data. To generate synthetic ground truth data, a single color magenta image is overlaid over a single color cyan image and then fused together with equal transparency. An example of synthetic image is shown in Figure \ref{fig:synthetic composite}. Ground truth masks for single color images are extracted by applying a simple thresholding method. Performance of the segmentation pipeline on synthetic ground truth data is excellent as shown in Table  \ref{table:accuracy}.

\begin{table}[h!]
\centering
\begin{tabular}{||c | c c||} 
 \hline
 Regions & mean IoU (\%) & pixel wise (\%) \\ [0.5ex] 
 \hline
 all regions & 95.39 & 92.54 \\
\hline
 cyan regions & 93.99 & 94.81  \\
 \hline
 magenta regions & 96.78 & 97.69 \\
\hline
\end{tabular}
\caption{Performance of the color segmentation pipeline. }
\label{table:accuracy}
\end{table}

\subsection{Ink-ink interactions and graininess noise}
\begin{figure}[h!]
    \centering
    \captionsetup{justification=centering}
        \begin{subfigure}[b]{0.09\textwidth}
            \centering
            \includegraphics[width=\textwidth]{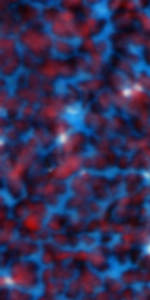}
            \caption[]
            {{\small Real}}
            \label{fig:real composite}
        \end{subfigure}
        \hfill
        \begin{subfigure}[b]{0.09\textwidth}
            \centering 
            \includegraphics[width=\textwidth]{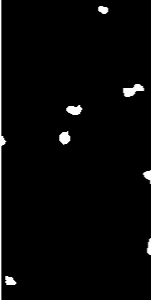}
            \caption[]%
            {{\small White}}    
            \label{fig:real b}
        \end{subfigure}
        \hfill
        \begin{subfigure}[b]{0.09\textwidth}
            \centering 
            \includegraphics[width=\textwidth]{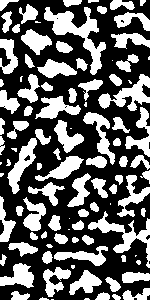}
            \caption[]%
            {{\small Cyan}}    
            \label{fig:real c}
        \end{subfigure}
        \hfill
        \begin{subfigure}[b]{0.09\textwidth}
            \centering 
            \includegraphics[width=\textwidth]{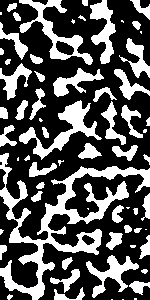}
            \caption[]%
            {{\small Magenta}}    
            \label{fig:real m}
        \end{subfigure}
        \hfill
        \begin{subfigure}[b]{0.09\textwidth}
            \centering 
            \includegraphics[width=\textwidth]{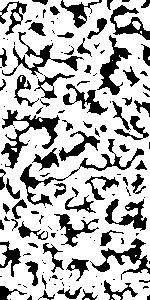}
            \caption[]%
            {{\small Overlap}}    
            \label{fig:real overlap}
        \end{subfigure}
        \vskip\baselineskip
       \begin{subfigure}[b]{0.09\textwidth}
            \centering
            \includegraphics[width=\textwidth]{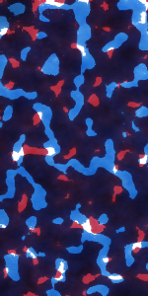}
            \caption[]%
            {{\small Synthetic}}
            \label{fig:synthetic composite}
        \end{subfigure}
        \hfill
        \begin{subfigure}[b]{0.09\textwidth}
            \centering 
            \includegraphics[width=\textwidth]{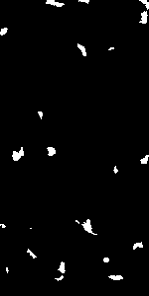}
            \caption[]%
            {{\small White}}    
            \label{fig:synthetic b}
        \end{subfigure}
        \hfill
        \begin{subfigure}[b]{0.09\textwidth}
            \centering 
            \includegraphics[width=\textwidth]{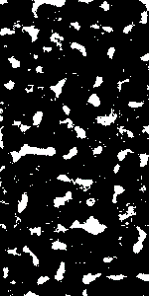}
            \caption[]%
            {{\small Cyan}}    
            \label{fig:synthetic c}
        \end{subfigure}
        \hfill
        \begin{subfigure}[b]{0.09\textwidth}
            \centering 
            \includegraphics[width=\textwidth]{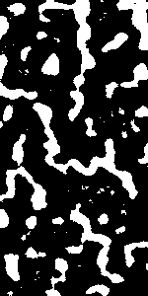}
            \caption[]%
            {{\small Magenta}}    
            \label{fig:synthetic m}
        \end{subfigure}
        \hfill
        \begin{subfigure}[b]{0.09\textwidth}
            \centering 
            \includegraphics[width=\textwidth]{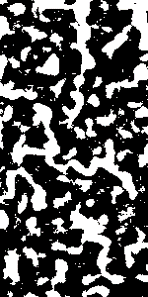}
            \caption[]%
            {{\small Overlap}}    
            \label{fig:synthetic overlap}
        \end{subfigure}
        \captionsetup{justification=raggedright}
        \caption{Comparison between real and synthetic superimposed image for 75\% Cyan and 75\% Magenta. In the masks, $black$ represent ink regions (of the corresponding color, or all colors for the white mask), while $white$ represents everything else. Note that the cyan and magenta masks includes regions of both pure color ($pc$ and $pm$) and overlap.}
        \label{fig: mask comparison}
\end{figure}

To investigate influences from ink-ink interaction on inkjet printing image quality, consider an experiment where real multi-color images are compared with synthetic superimposed images. In a synthetic superimposed image, a magenta-only printed image is superimposed on a cyan-only printed image, with ink levels same as the corresponding real multi-color image. In the synthetic superimposed image, there is no interaction between cyan and magenta inks. But in the real image, the cyan ink is laid down first and then the magenta ink, and these ink layers can flow and interact with each other over time scales of 10-100 milliseconds until the image stabilizes.  Figures \ref{fig:real composite}
 and \ref{fig:synthetic composite}  shows an example of the real image and synthetic superimposed image, respectively,  and their corresponding masks in \ref{fig:real b}-\ref{fig:real overlap} and \ref{fig:synthetic b}-\ref{fig:synthetic overlap}, respectively.

A comparison between the masks of the real and superimposed synthetic images shows how ink-ink interactions are disrupting the ink patterns. No ink regions ($white$ regions in the white mask) on the real image are much larger and more sparsely distributed than those in the synthetic image. In the cyan and magenta masks, ink patterns (black regions) in real images are finer and more fragmented. The fragmentation of ink patterns is also evident in the overlap mask for the real image. This suggests that cyan and magenta inks try to avoid each other, leading to much less coalescence than one would expect based purely on ink levels. 

To investigate the connection between segmentation color components and visually perceptible graininess noise, consider the correlation between band-pass filtered reflectance of original image and band-pass filtered segmentation masks of each color components is considered.  A $Butter$-$worth$ band-pass filter \cite{butterworth1930theory} is used with cut-off frequencies $1$ to $10$ $cycles/mm$, to filter out all other noise except for graininess, including low frequency noise (e.g. mottle < 1 cycle/mm) and very high frequency noise (e.g. halftone dot pattern > 10 cycles/mm).  This is a reasonable approach because the reflectance of an image is directly related to its luminescence(i.e. L*), and the band-passed reflectance image is equivalent to graininess noise which is often characterized by variations in L* between frequencies of 1-10 cycles/mm.   

    

\begin{figure}[h]
    \centering
    \captionsetup{justification=centering}
        \begin{subfigure}[b]{0.234\textwidth}
            \centering
            \includegraphics[width=\textwidth]{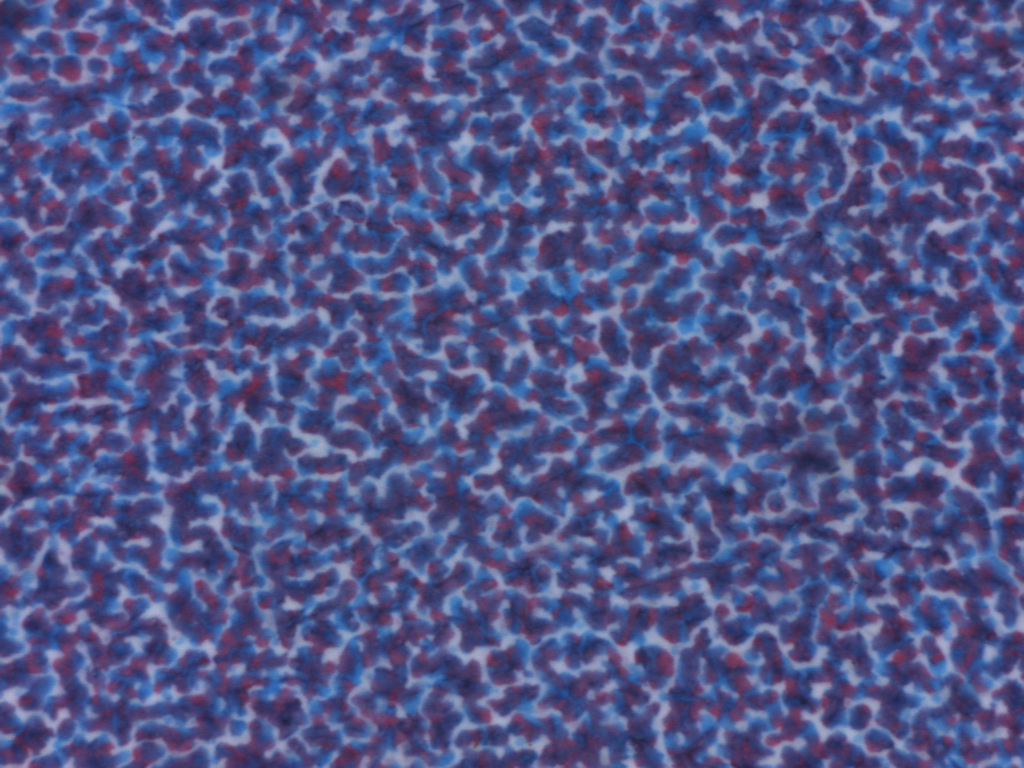}
            \caption[Network2]%
            {{\small Original image.}}
            \label{fig:origin_img}
        \end{subfigure}
        \hfill
        \begin{subfigure}[b]{0.234\textwidth}
            \centering 
            \includegraphics[width=\textwidth]{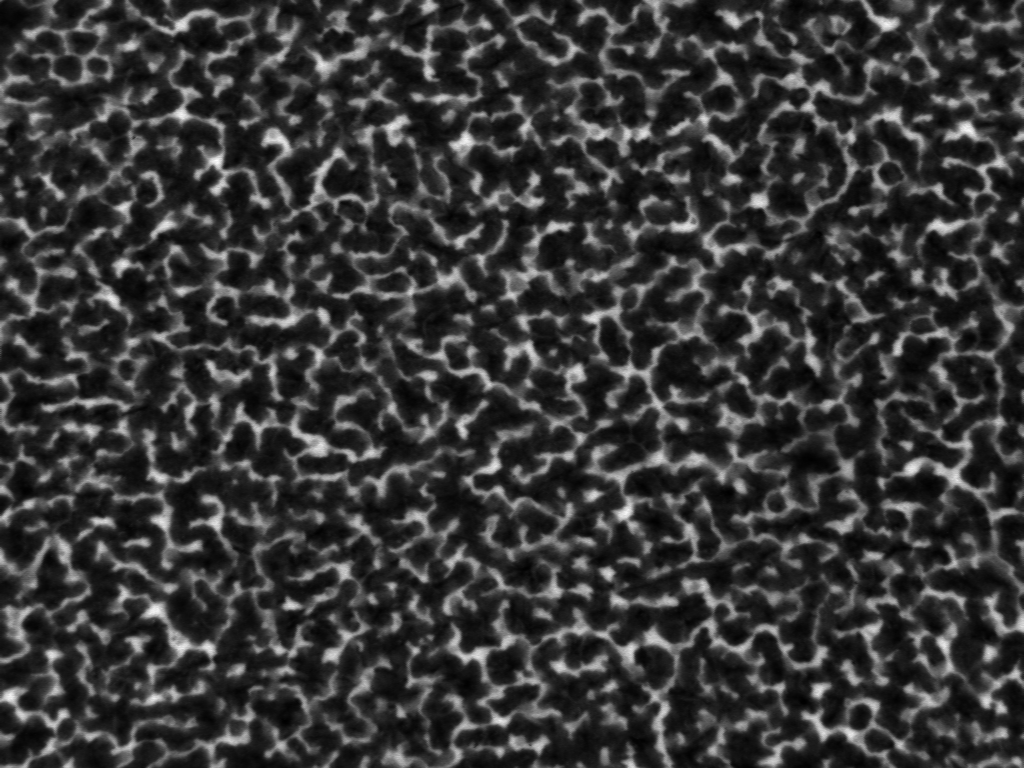}
            \caption[Nerwork2]%
            {{\small Original reflectance image.}}    
            \label{fig:origin_ref}
        \end{subfigure}
        \vskip\baselineskip
        \begin{subfigure}[b]{0.234\textwidth}
            \centering
            \includegraphics[width=\textwidth]{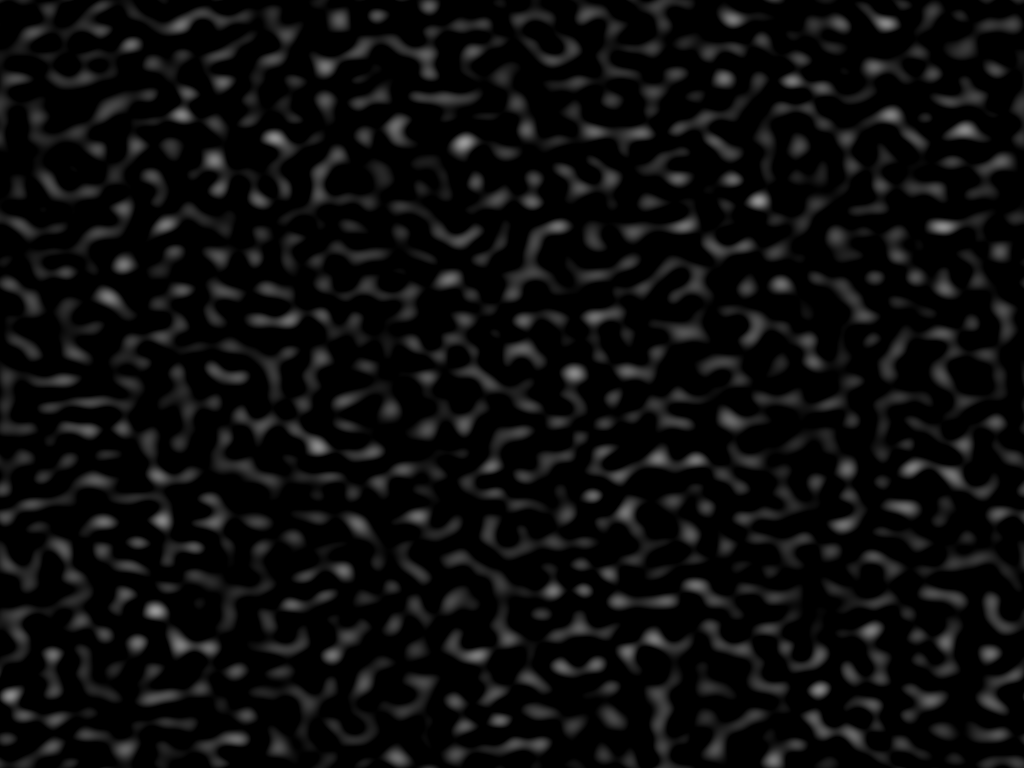}
            \caption[Network2]%
            {{\small Band-pass filtered original reflectance image.}}
            \label{fig:origin_bp}
        \end{subfigure}
        \hfill
        \begin{subfigure}[b]{0.234\textwidth}
            \centering 
            \includegraphics[width=\textwidth]{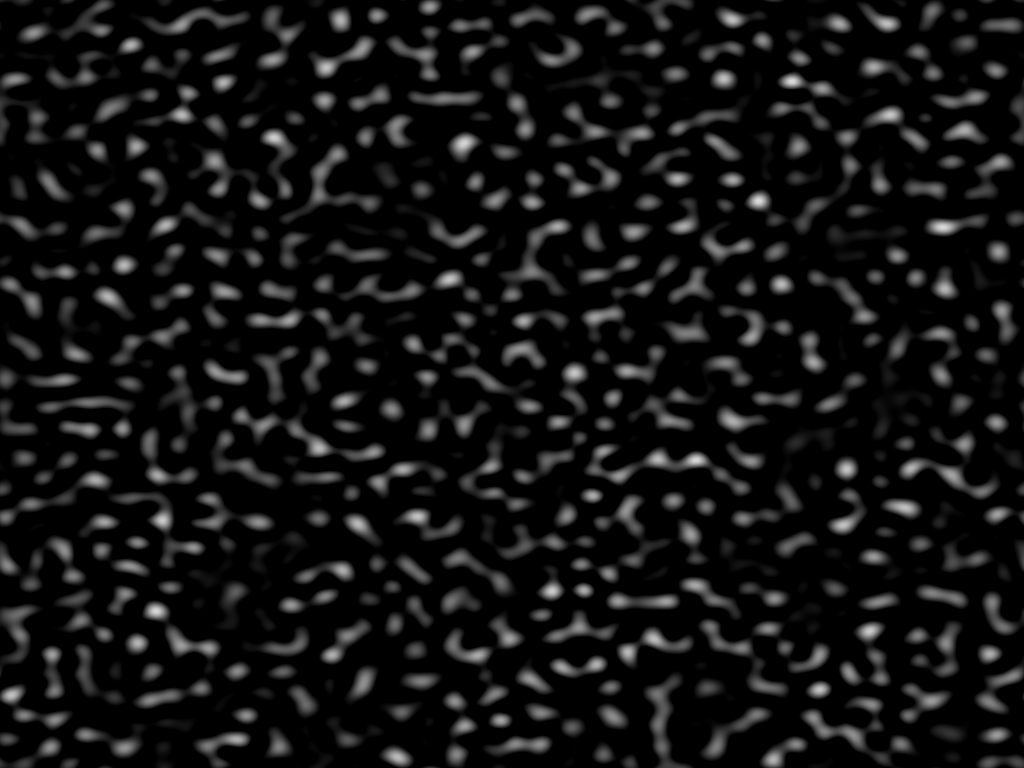}
            \caption[Network2]%
            {{\small Band-pass filtered reconstructed reflectance image.}}    
            \label{fig:syn_bp}
        \end{subfigure}
        \caption{Original and band-pass filtered reflectance images for Cyan 30\%, Magenta 30\%.}
        \label{fig:bandpassfilter}
\end{figure}

Figures \ref{fig:origin_bp} and \ref{fig:syn_bp} shows an example of the band-pass filtered reflectance of original image and band-pass filtered reconstructed reflectance image (Equation \ref{eq:our model}). The similarity between the two images is quite apparent (correlation coefficient is $0.96$). This proves that the validity of the reflectance model (Equation \ref{eq:our model}), especially in the frequency range of interest.  Table \ref{table:corr_coeff} quantifies the contribution of each color component to image graininess by correlating the band-pass filtered segmentation mask and the band-pass filtered reflectance image. It can be observed that the white mask has the highest correlation, indicating that the graininess noise in multi-color inkjet images is largely due to the no ink regions in the image.

\begin{table}[h!]
\centering
\begin{tabular}{||c | c | c | c ||} 
\hline
 Color Component & Mean Correlation Coefficient \\
\hline
Pure Cyan & 0.38577938 \\
 Pure Magenta & 0.11904254 \\
 Overlap & 0.02307696 \\
 Blank & 0.872573 \\
\hline
\end{tabular}
\caption{Mean Correlation coefficients between band-pass filtered color components and band-pass filtered original reflectance images across the dataset. }
\label{table:corr_coeff}
\end{table}
\section{Conclusion}
In this paper, a systematic approach to study the ink pattern distributions in multi-color inkjet printed images outlined. A color segmentation pipeline is developed to decompose high resolution inkjet printed images into independent color components. The color segmentation pipeline is used to study ink-ink interactions in multi-color inkjet printed images. A reflectance model is proposed to relate these color components to the reflectance of the overall image, and is used investigate the role of different color components to graininess. 

\bibliographystyle{IEEEbib}
\bibliography{ms}

\begin{thebibliography}{10}

\bibitem{rasmussen2001measurement}
Rene Rasmussen, Edul~N Dalal, and Kristen Hoffman,
\newblock ``Measurement of macro-uniformity: Streaks, bands, mottle and
  chromatic variations,''
\newblock in {\em PICS}, 2001, pp. 90--95.

\bibitem{dalal1998evaluating}
Edul~N Dalal, D~Ren{\'e} Rasmussen, Fumio Nakaya, Peter~A Crean, and Masaaki
  Sato,
\newblock ``Evaluating the overall image quality of hardcopy output,''
\newblock in {\em PICS}, 1998, pp. 169--173.

\bibitem{rasmussen2006iso}
D~Ren{\'e} Rasmussen, Kevin~D Donohue, Yee~S Ng, William~C Kress, Frans
  Gaykema, and Susan Zoltner,
\newblock ``Iso 19751 macro-uniformity,''
\newblock in {\em Image Quality and System Performance III}. International
  Society for Optics and Photonics, 2006, vol. 6059, p. 60590K.

\bibitem{wang2013figure}
Weibao Wang, Gary Overall, Travis Riggs, Rebecca Silveston-Keith, Julie
  Whitney, George Chiu, and Jan~P. Allebach,
\newblock ``Figure of merit for macrouniformity based on image quality ruler
  evaluation and machine learning framework,''
\newblock in {\em Image Quality and System Performance X}. International
  Society for Optics and Photonics, 2013, vol. 8653, p. 86530P.

\bibitem{chen2019segmentation}
Qiulin Chen, Renee Jessome, Eric Maggard, and Jan~P. Allebach,
\newblock ``Segmentation-based detection of local defects on printed pages,''
\newblock {\em Electronic Imaging}, vol. 2019, no. 10, pp. 301--1, 2019.

\bibitem{huang2019cost}
Wan-Eih Huang, Eric Maggard, Renee Jessome, Yousun Bang, Minki Cho, and Jan~P.
  Allebach,
\newblock ``Cost-function-based repetitive interval estimation method with
  synthetic missing bands for periodic bands in electrophotographic printer,''
\newblock {\em Electronic Imaging}, vol. 2019, no. 10, pp. 302--1, 2019.

\bibitem{jing2013general}
Xiaochen Jing, Steve Astling, Renee Jessome, Eric Maggard, Terry Nelson, Mark
  Shaw, and Jan~P Allebach,
\newblock ``A general approach for assessment of print quality,''
\newblock in {\em Image Quality and System Performance X}. International
  Society for Optics and Photonics, 2013, vol. 8653, p. 86530L.

\bibitem{xiang2019blockwise}
Xiaoyu Xiang, Renee Jessome, Eric Maggard, Yousun Bang, Minki Cho, and Jan~P.
  Allebach,
\newblock ``Blockwise based detection of local defects,''
\newblock {\em arXiv preprint arXiv:1906.02374}, 2019.

\bibitem{yan2015autonomous}
Ni~Yan, Eric Maggard, Roberta Fothergill, Renee~J Jessome, and Jan~P Allebach,
\newblock ``Autonomous detection of iso fade point with color laser printers,''
\newblock in {\em Image Quality and System Performance XII}. International
  Society for Optics and Photonics, 2015, vol. 9396, p. 93960F.

\bibitem{zhang2019block}
Runzhe Zhang, Eric Maggard, Renee Jessome, Yousun Bang, Minki Cho, and Jan~P.
  Allebach,
\newblock ``Block window method with logistic regression algorithm for streak
  detection,''
\newblock {\em Electronic Imaging}, vol. 2019, no. 10, pp. 300--1, 2019.

\bibitem{nguyen2014perceptual}
Minh~Q Nguyen, Renee Jessome, Steve Astling, Eric Maggard, Terry Nelson, Mark
  Shaw, and Jan~P Allebach,
\newblock ``Perceptual metrics and visualization tools for evaluation of page
  uniformity,''
\newblock in {\em Image Quality and System Performance XI}. International
  Society for Optics and Photonics, 2014, vol. 9016, p. 901608.

\bibitem{hertel2003one}
Dirk~W Hertel and Bror~O Hultgren,
\newblock ``One-step measurement of granularity versus density, graininess, and
  micro-uniformity,''
\newblock in {\em IS\&T's PICS Conference}. Society for Imaging Science \&
  Technology, 2003, pp. 552--557.

\bibitem{tse2009towards}
Ming-Kai Tse and David Forrest,
\newblock ``Towards instrumental analysis of perceptual image and print
  quality,''
\newblock in {\em NIP \& Digital Fabrication Conference}. Society for Imaging
  Science and Technology, 2009, vol. 2009, pp. 528--531.

\bibitem{otsu1979threshold}
Nobuyuki Otsu,
\newblock ``A threshold selection method from gray-level histograms,''
\newblock {\em IEEE transactions on systems, man, and cybernetics}, vol. 9, no.
  1, pp. 62--66, 1979.

\bibitem{cover1967nearest}
Thomas Cover and Peter Hart,
\newblock ``Nearest neighbor pattern classification,''
\newblock {\em IEEE transactions on information theory}, vol. 13, no. 1, pp.
  21--27, 1967.

\bibitem{berns2000principles}
S~Berns~Roy,
\newblock ``Principles of color technology,'' 2000.

\bibitem{tse2007pias}
Ming-Kai Tse,
\newblock ``{PIAS-II}\texttrademark--{A} high-performance portable tool for
  print quality analysis anytime, anywhere,''
\newblock {\em Journal of the Imaging Society of Japan, Tokyo, Japan}, pp.
  1--4, 2007.

\bibitem{butterworth1930theory}
Stephen Butterworth et~al.,
\newblock ``On the theory of filter amplifiers,''
\newblock {\em Wireless Engineer}, vol. 7, no. 6, pp. 536--541, 1930.

\end{thebibliography}


\end{document}